\documentclass[fleqn,usenatbib]{mnras}

\usepackage{newtxtext,newtxmath}
\usepackage[dvipsnames]{xcolor}

\usepackage[T1]{fontenc}
\usepackage{lineno}

\DeclareRobustCommand{\VAN}[3]{#2}
\let\VANthebibliography\thebibliography
\def\thebibliography{\DeclareRobustCommand{\VAN}[3]{##3}\VANthebibliography}

\usepackage{graphicx}	
\usepackage{amsmath}	
\usepackage{amssymb}	
\usepackage{tabularx}
\usepackage[skip=0pt]{caption}
\usepackage{color,soul}

\title[Disc-Jet Decomposition of Fermi Blazars ]{Accretion Disc-Jet Decomposition from the Optical-Near Infrared Monitoring of Fermi Blazars}

\author[G. Rajguru et al.]{
Garima Rajguru$^{1,2}$\thanks{E-mail: cosmogarima@gmail.com} and
Ritaban Chatterjee$^{3}$
\\
$^{1}$Department of Physics and Astronomy, Clemson University, 105 Sikes Hall, Clemson, SC 29634, USA.\\
$^{2}$Department of Physics, Presidency University, 86/1 College Street, Kolkata-700073, WB, India.\\
$^{3}$ School of Astrophysics, Presidency University, 86/1 College Street, Kolkata-700073, WB, India.
}

\date{Accepted XXX. Received YYY; in original form ZZZ}

\pubyear{2015}

\begin{document}

\label{firstpage}
\pagerange{\pageref{firstpage}--\pageref{lastpage}}
\maketitle

\newcommand{\bbe}{\begin{eqnarray}}
\newcommand{\ee}{\end{eqnarray}}
\newcommand{\n}{\nonumber}

\begin{abstract}
We study the variability of the thermal (accretion disc) and non-thermal (jet) emission of thirteen flat spectrum radio quasars in the optical and near infrared (OIR) regime using light curves spanning years with an average sampling of three observations per week. We fit a combination of a blackbody and a power-law function to the OIR data, in the blazar rest frame, to extract the corresponding thermal (disc) and non-thermal (jet) components from the total flux. We carry out this analysis for the entire duration of the light curves to obtain the variation of the disc and jet components over years. Reliability of our fits have been affirmed by successfully retrieving accurate parameters by employing our method to simulated data and by comparing our results with published disc luminosity obtained by other methods for a few well-observed blazars. In blazars, the thermal (disc) emission is difficult to extract because the relativistically beamed radiation of the jet dominates at all wavelengths. By employing this method, the disc emission in blazars may be estimated directly from photometric data at OIR bands instead of indirect methods, such as, inferring it from the emission line luminosities. We find that the variability of the disc and jet emission obtained by the above method are strongly correlated in most cases. 
\end{abstract}


\section{Introduction}
Certain properties of various types of active galactic nuclei (AGN) are attributed to their orientation with respect to our line of sight, according to the unified scheme of AGN \citep{1995PASP..107..803U}. Blazars are a class of AGN, which possess energetic jets pointed almost directly towards the Earth with an angle between the jet axis and our line of sight $\theta <10^\circ$.
The relativistic electrons present in the jet give rise to radio to optical (sometimes X-ray) emission through synchrotron radiation \citep{1981Natur.293..714B,1982ApJ...253...38U,1988AJ.....95..307I,1998ASPC..144...25M}. In the so-called ``leptonic model of blazar emission,'' inverse-Compton effect is presumed to be the origin of the high energy (keV$-$GeV) emission, in which the same electrons up-scatter the jet synchrotron photons \citep{1992ApJ...397L...5M, 2002ApJ...564...92C, 2005ApJ...627...62A} or external photons, from the broad line region (BLR) or dusty torus, to X-rays and $\gamma$-rays \citep{1994ApJ...421..153S, 1999ApJ...521L..33C, 2000ApJ...545..107B, 2009ApJ...692...32D}. 
 The so-called `big blue bump' in the observed spectra of AGN may be crudely identified with that from a blackbody \citep{fkr_book2002}. On the other hand, the accretion disc is modeled as optically thick material, the effective temperature of which decreases with radius and has a maximum of $\sim 10^4-10^5$ K at the center \citep{1973A&A....24..337S}. Therefore, the thermal emission in the spectra of AGN at the UV-optical-NIR wavelengths is considered to be generated in the disc \citep{malkan&sargent1982,1978Natur.272..706S}. Indeed, several works indicate that the optical emission in AGN is governed by disc emission, such as, from the Balmer edge absorption observed in the optical spectra in polarized light \citep{2003MNRAS.345..253K} and the study of the optical color-luminosity relations in AGN \citep{2005MNRAS.363...57L}, among others. We note, however, that some observations, particularly in the UV band, have been found to disagree with the simple accretion disk model \citep[e.g.,][]{2012MNRAS.423..451L}. 
\par

Based on the equivalent width (EW) of their broad emission lines in the rest frame, blazars may be categorized into two major groups. Blazars which are supposed to be devoid of prominent emission lines (EW $<5$\AA), presumably due to the presence of a weak accretion disc, are called BL Lacertae (BL Lacs) objects, whereas those that possess prominent emission lines (EW $>5$\AA) in their spectra \citep{1995PASP..107..803U} are termed as flat spectrum radio quasars (FSRQ). However, this segregation may not be robust \citep{2011MNRAS.414.2674G}, since bright emission lines can be detected sporadically in some BL Lacs. As more precise observations have been available in recent years \citep{1995ApJ...452L...5V,2010A&A...516A..59C} some BL Lacs have been known to exhibit emission lines with EW $>5$\AA~. Nevertheless, FSRQs are the suitable targets for studying the disc emission in blazars. 
\par

The launching and collimation of jets are not well understood. The extraction of rotational energy of the super-massive black hole (SMBH) at the center \citep{1977MNRAS.179..433B} or extraction of power from the accretion disc in the presence of magnetic fields sustained by the disc \citep{1982MNRAS.199..883B}, are two accepted theories. Mechanism of jet launching and collimation is, thus, assumed to be related to the accreting matter and magnetic fields associated with the disc \citep{lind&meier1989,mckinney&narayan2007,Tchekhovskoy&narayan2011,2013ApJ...770...31W}. Hence, a relation between the dynamics of the disc and the jet, and their properties is anticipated. 

\par
Blazars have a prominent jet and are suitable sources for investigating the launching and collimation of jets from near a SMBH. On the other hand, since the blazar jet emission is relativistically beamed in the observer's frame it outshines the rest of the AGN, e.g., accretion disc, BLR and torus. As a result, studying the accretion disc properties of blazars and its relation to the jet emission is challenging. \citet{2022PhRvD.106f3001R} studied the disc and jet power of a large sample of Fermi blazars in order to investigate any relation between those properties. In that paper, emission line luminosities were used as indicators of the disc emission. Here, we analyze the optical-near infrared (OIR) observations of a number of FSRQs at $BVRJK$ bands to decompose the total flux into disc and jet components in order to obtain an estimate of the disc emission directly. FSRQs are known to show signatures of thermal emission in their spectra \citep{1997ApJ...484..108S, 1988ApJ...326L..39S, 1991ApJ...375...46I, 1989ApJ...340..129B, 1992ApJ...398..454W, 1989ApJ...340..150B}. Since the jet is usually dominant and variable, it is in the low state of jet activity which we concentrate on to detect the thermal component. When the jet is in a low state of its variable luminosity, excess emission may be observed at the optical-UV range in addition to the underlying power-law spectrum of the jet, which may be due to thermal emission from the accretion disc. We find the properties of the excess emission, and assuming it exhibits a blackbody spectrum arising from the accretion disc, we deduce the properties of the latter. Furthermore, we carry out the above decomposition for $BVRJK$ monitoring of several blazars over eight years to study the variation in the accretion disc properties and its relation with the jet. 
\begin{figure}
\centering
\includegraphics[width=8cm,height=8cm]{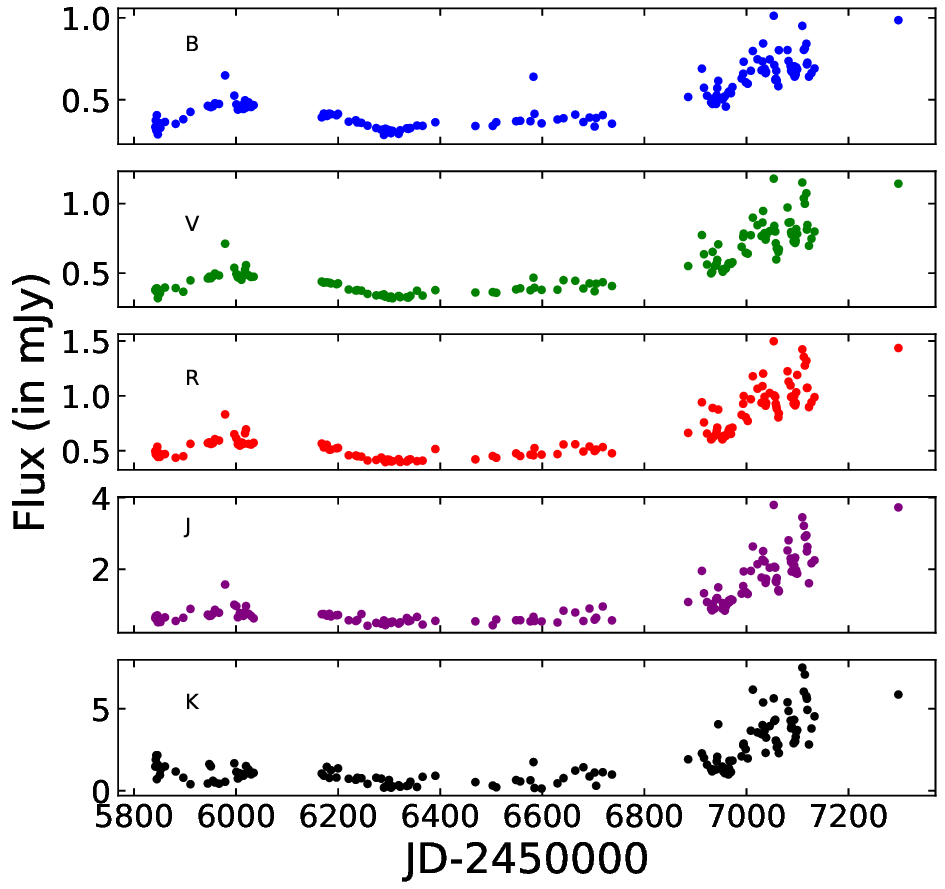}
\caption{SMARTS optical and near IR light curves of PKS 0402-362.}
\label{fig1}
\end{figure}

\par
In this paper, Section \ref{sec2} deals with sample selection and data obtained for our sources. Section \ref{sec3} elucidates the analysis done to decompose the OIR fluxes into the thermal and jet components. In this section, we further check the efficacy of the method we have employed using simulated data and by applying our method to a few sources for which more extensive data are available and comparing its outcome with published results. Section \ref{sec4} provides the discrete cross correlation function computed for the decomposed thermal (disc) and non-thermal (jet) luminosities in our blazar sample. Section \ref{sec5} presents the conclusion of our analysis.


\section{Data and Sample Selection}
\label{sec2}
Optical and near IR photometric data are available for a vast collection of blazars from the Yale-SMARTS blazar monitoring project{\footnote{\href{http://www.astro.yale.edu/smarts/}{http://www.astro.yale.edu/smarts/}}}  \citep{2012ApJ...756...13B,2012ApJ...749..191C}. As a part of this project, blazars, which are observable from the southern hemisphere and were detected by Fermi-LAT were monitored with a variable cadence during 2008-2015. Blazars, which were flaring in the GeV band were observed more frequently and \textit{vice versa}. The 1.3 m telescope, which is a part of the Small and Moderate Aperture Research Telescope System (SMARTS) consortium that operates a number of telescopes at Cerro Tololo, Chile, was used for the above observations. The data were taken by the ANDICAM instrument, which is able to carry out photometric observations at one optical and one near IR band simultaneously \citep{2003SPIE.4841..827D}. $BVRJK$-band light curves of $\sim$100 blazars with a range of total duration and cadence were obtained in the Yale-SMARTS blazar monitoring project.

In this work, we select 13 blazars, for which the length and average cadence of the $BVRJK$ light curves are at least 2 years and three observations per week, respectively. Figure \ref{fig1} shows such a set of $BVRJK$ light curves of the blazar PKS 0402-362. 
Light curves with a high cadence is necessary to probe the detailed properties of the different components of the variable emission and its long-term trend may be studied using long trains of light curves. The 13 blazars of our sample, all belonging to the FSRQ subclass, are shown in Table \ref{Tab0} along with their redshifts, taken from the recently published $Fermi$ Gamma Ray Space Telescope Fourth Catalog of Active Galactic Nuclei detected by the LAT (4LAC-DR3, \citet{4lac}). 

\begin{table}
	\caption{The blazar (FSRQ) sample and their redshifts.}
	\begin{tabularx}{\columnwidth}{|X|X|}
	\hline
	Blazar Name & Redshift ($z$)\\
	\hline
	\hline
	PKS 0208-512 & 1.003\\
        \hline
        PKS 0402-362 & 1.417\\
        \hline
        PKS 0454-234 & 1.003\\
        \hline
        3C 273 & 0.158\\
        \hline
        3C 279 & 0.536\\
        \hline
        PKS 1406-076 & 1.493\\
        \hline
        PKS 1424-41 & 1.522\\
        \hline
        PKS 1510-089 & 0.36\\
        \hline
        PKS 1622-297 & 0.815\\
        \hline
        PKS 1730-130 & 0.902\\
        \hline
        PKS 2052-474 & 1.489\\
        \hline
        PKS 2142-75 & 1.139\\
        \hline
        3C 454.3 & 0.859\\
	\hline
	\end{tabularx}
\label{Tab0}
\end{table}

The fluxes were corrected for Galactic extinction using the data from the NASA/IPAC Extragalactic Database (NED) {\footnote{\href{https://ned.ipac.caltech.edu/}{https://ned.ipac.caltech.edu/}}}. UV data points, in a few cases, were collected from the Swift-UV-Optical Telescope (UVOT), publicly available at Space Science Data Center (SSDC) website\footnote{\href{https://www.ssdc.asi.it/mmia/index.php?mission=swiftmastr}{https://www.ssdc.asi.it/mmia/index.php?mission=swiftmastr}}. The optical luminosity of blazar host galaxies, at these redshifts, is negligible compared to the nuclear luminosity \citep{Sbarufatti2005, Shaw2013, Goldoni2021}. Therefore, the host galaxy contribution in the blazars in our sample, in images taken by the 1.3 m telescope, have been ignored  \citep{2012MNRAS.420.2899G,2012MNRAS.425.3002G, Landoni_2015, urry2000hubble, 1993A&AS...98..393S}. 


\section{Analyses}
\label{sec3}
\subsection{Disc-Jet Decomposition Method}
For each blazar in our sample, the $BVRJK$-band fluxes were fit with a combination of  thermal blackbody and non-thermal power-law components:
\bbe
\label{eqn1}
f_\nu &=& a\nu^{-b} + \frac{c_b\nu^3}{e^{d\nu}-1} 
\ee
where $\nu$ is the rest-frame frequency, $a$ and $c_b$ are normalization constants, $b$ is the spectral index, $d$ is a parameter related to the temperature of the blackbody. The power-law corresponds to the non-thermal synchrotron emission of the jet while the thermal blackbody is an approximate representation of the accretion disc emission.
Hence, this enables us to decompose the total flux into a disc and a jet component. We obtain the spectral index of the blazar emission and effective temperature of the accretion disc from the best-fit values of the free parameters. Temperature is calculated from the free parameter $d$ by fitting the curve at a low state of the jet, when the thermal component is prominent. We repeat the same analysis for each day of the years-long light curve to obtain the variability of the best-fit disc and jet components over the duration of the light curve.
Thus, we can calculate the day-to-day flux variation of the power-law (PL) and blackbody (BB) components using Eqn. \ref{eqn1}

\begin{figure*}
\centering
\includegraphics[width=\textwidth]{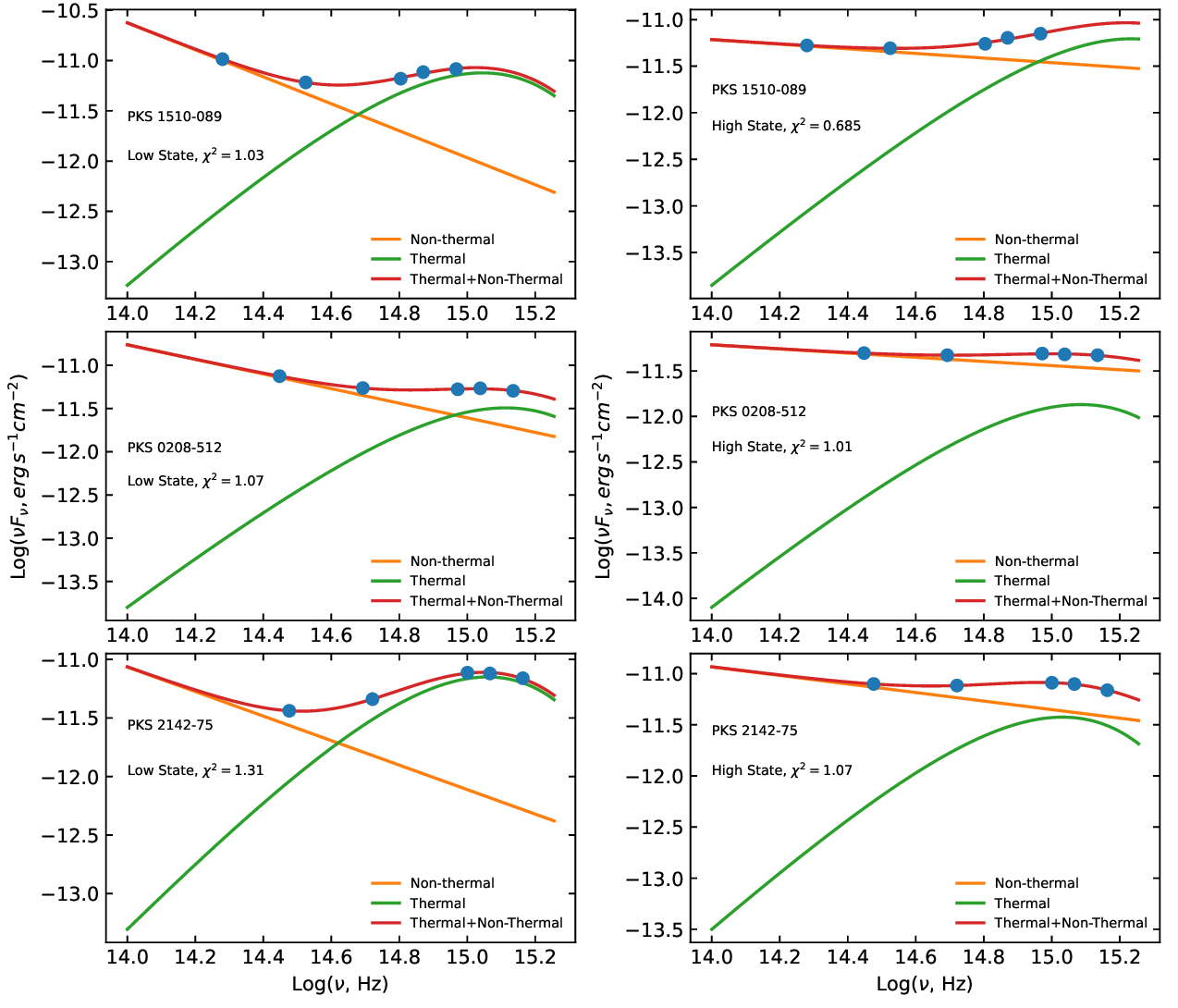}
\caption{ Each panel shows an example of a best-fit to the OIR data points at a high or low state  of a blazar using our model, in the blazar rest-frame. \textbf{Top Left:} Low state of PKS 1510-089. \textbf{Top Right:} High state of PKS 1510-089.  \textbf{Middle Left:} Low state of PKS 0208-512. \textbf{Middle Right:} High state of PKS 0208-512. \textbf{Bottom Left:} Low state of PKS 2142-75. \textbf{Bottom Right:} High state of PKS 2142-75. The reduced-$\chi^2$ values are indicated in the plots.}
\label{HighLow}
\end{figure*}

\begin{figure*}
\centering
\includegraphics[width=\textwidth]{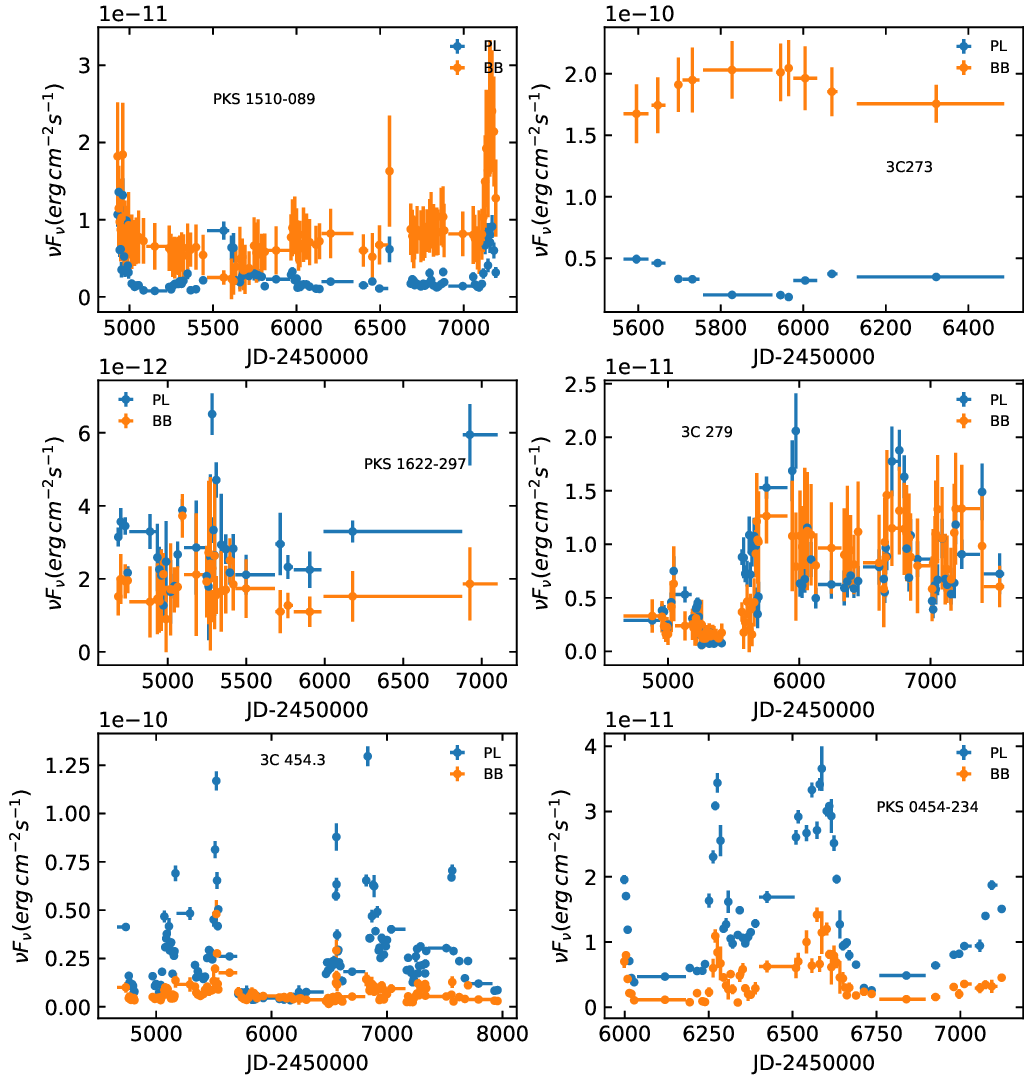}
\caption{Each panel shows binned PL and BB components vs time plots of \textbf{Top Left:} PKS 1510-089. \textbf{Top Right:} 3C 273.  \textbf{Middle Left:} PKS 1622-297. \textbf{Middle Right:} 3C 279. \textbf{Bottom Left:} 3C 454.3. \textbf{Bottom Right:} PKS 0454-234.}
\label{Bin}
\end{figure*}
\par
Plugging the best-fit parameters in the first and second terms on the right-hand side of Eqn. \ref{eqn1} respectively, we can compute the decomposed PL and BB flux at each frequency. The disc and jet flux has been calculated at a frequency of $\nu=6.81\times 10^{14}$ Hz in the blazar rest frame. Fig. \ref{HighLow} shows the OIR spectra of the high and low states of a few of the blazars in our sample. The low state of a blazar is characterised by the relative increase in thermal (blackbody) contribution to the overall SED, compared to a high state where the non-thermal (power law) component dominates. Moreover, the high state has comparatively higher flux than a low state due to increase in jet emission. Before proceeding to apply the above analysis to the detailed light curves, we bin the light curves over longer time intervals and decompose the average flux of the bins into a disc and a jet component. The binning has been done over intervals of 5 days of observation. We plot the PL and BB components of each of the bins in order to demonstrate the average behaviour of high and low states over time (Figure \ref{Bin}). Errors obtained on $\chi^2$ fitting have been used to estimate the errors on the binned fluxes.

\begin{figure*}
\centering
\includegraphics[width=\textwidth]{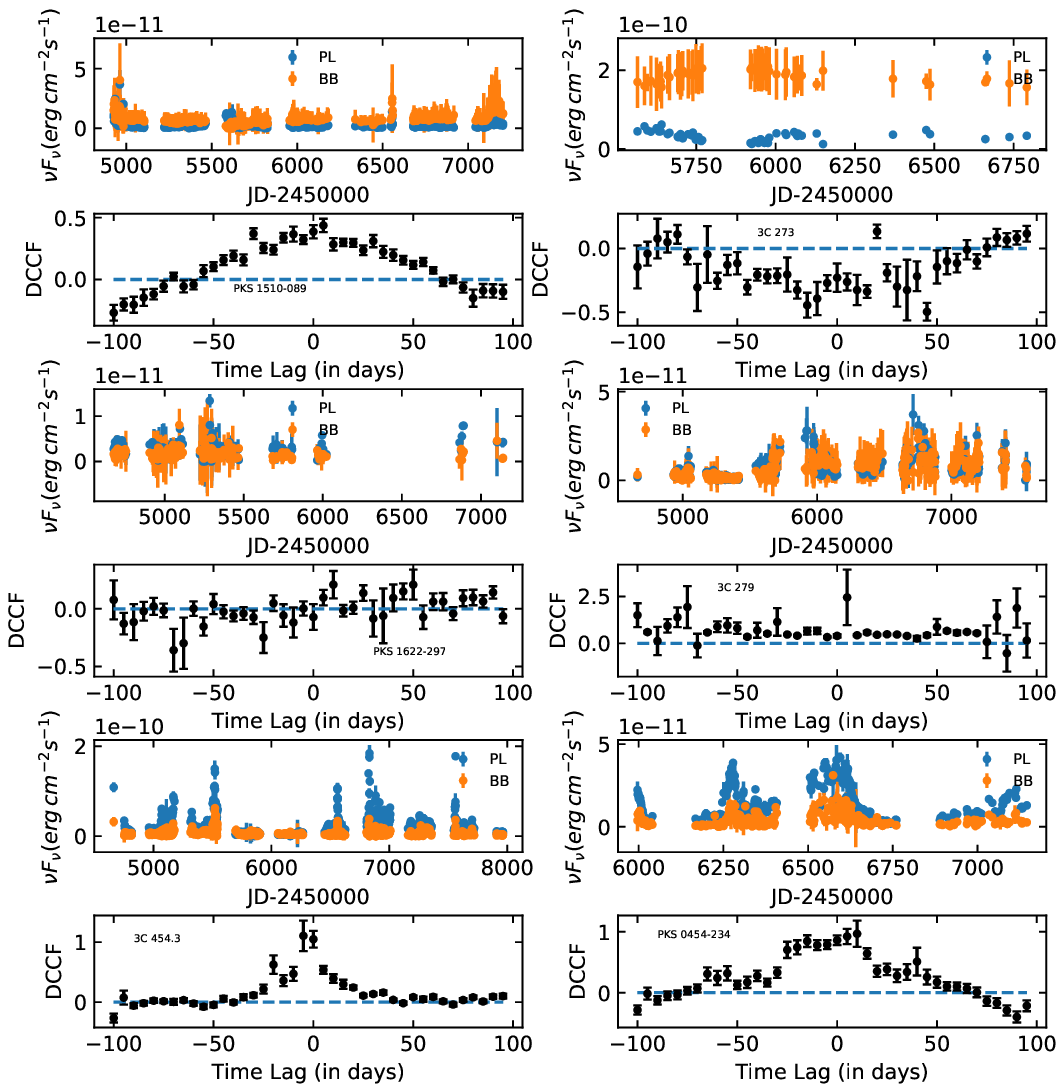}
\caption{Each panel shows the decomposed disc and jet component light curves and the associated DCCF plot below it. \textbf{Top Left:} PKS 1510-089. \textbf{Top Right:} 3C 273.  \textbf{Middle Left:} PKS 1622-297. \textbf{Middle Right:} 3C 279. \textbf{Bottom Left:} 3C 454.3. \textbf{Bottom Right:} PKS 0454-234.}
\label{DCCF}
\end{figure*}

\subsection{Results}
Plots corresponding to the disc-jet decomposition in individual epochs are shown in 
Figure \ref{HighLow} for the blazars PKS 1510-089, PKS 0208-512 and PKS 2142-75. The best-fit function containing thermal (BB) and non-thermal (PL) components are shown, for both high and low states. Example fits of the remaining blazars are given in Appendix A: Figure \ref{Fit_Remaining}.

\par
For high states, the non-thermal component dominates the thermal component, over the frequency range of observation. For low states, the contribution of the thermal component is significant and a peak in the thermal blackbody function near the optical-UV region is observed. This agrees with several studies which found the thermal bump peaking near UV, thereby suggesting temperatures of the order $10^4$ K \citep[e.g.,][]{1974ApJ...191..507T,1973A&A....24..337S}. Table \ref{Tab1} gives the estimated parameters for the 3 blazars in low states. From Table \ref{Tab1}, we see that our results reconcile with previous studies.\\ 
We compare the parameters estimated from our analysis of the OIR data, to those obtained from multiwavelength SED modelling found in literature, to check for consistency. Table \ref{TabTemp} compares the estimated temperatures obtained from our analysis to that obtained from  \citet{krauss2016tanami}. We see that the results are consistent. Table \ref{TabDiscLum} compares the estimated disc luminosities obtained from our analysis to that obtained from  \citet{ghisellini2010general}. We note that the disc luminosities found in literature (e.g., from Ghisellini et al. 2010) obtained using SED modelling have a large uncertainty ($\sim$50\%) due to parameter degeneracy as well as variability. Taking  the uncertainty into account, our estimates of disc luminosity agree with the values obtained from literature. Thus, we see that a simple blackbody fits the data just as well as a multicolor blackbody. This is consistent with the findings of \citet{krauss2016tanami}. They studied the SED of 22 radio-loud AGN at multiple epochs. In all the 81 SEDs they analysed, they found that a single-temperature blackbody fit the thermal component satisfactorily and better than a multicolor blackbody.

\begin{table*}
	\caption{Parameters estimated by applying our method to the low state of three blazars}
	\begin{tabularx}{\textwidth}{|X|X|X|X|X|X|}
	\hline
	Blazar & a & b & $c_b$& d & Temperature(in K)\\
	\hline
	\hline
	PKS 1510-089  & 23.72 ($\pm$1.42) & 2.34 ($\pm$0.09) & 0.024 ($\pm$0.008) & 0.353 ($\pm$0.035) & 13579 ($\pm$1347)\\
	\hline
	PKS 0208-512 & 17.25 ($\pm$0.68) & 1.84 ($\pm$0.04) & 0.005 ($\pm$0.001) & 0.302 ($\pm$0.012) & 15879 ($\pm$676)\\
	\hline
	PKS 2142-75 & 8.60 ($\pm$2.97) & 2.04 ($\pm$0.34) & 0.020 ($\pm$0.003) & 0.341 ($\pm$0.010) & 14049 ($\pm$436)\\
	\hline
	\end{tabularx}
\label{Tab1}
\end{table*}

\begin{table*}
	\caption{Comparison of disc temperatures given in the literature \citep{krauss2016tanami} with that estimated from our analysis.}
	\begin{tabularx}{\textwidth}{|X|X|X|}
	\hline
	Blazar & Temperature ( $\times 10^4$ K) from literature & Estimated Temperature ( $\times 10^4$ K)\\
	\hline
	\hline
        PKS 0208-512 & 1.35-1.49 & 1.58 ($\pm$0.06)\\
        \hline
        PKS 0402-362 & 1.3-1.4 & 1.64 ($\pm$0.37)\\
        \hline
        PKS 1424-41 & 1.18-1.31 & 1.22 ($\pm$0.06)\\
	\hline
        PKS 2052-474 & 1.25 & 1.23 ($\pm$0.33)\\
	\hline
        PKS 2142-75 & 1.3-1.4 & 1.40 ($\pm$0.04)\\
        \hline
	\end{tabularx}
\label{TabTemp}
\end{table*}

\begin{table*}
	\caption{Comparison of disc luminosity estimated from our analysis with that given in the literature \citep{ghisellini2010general}.}
	\begin{tabularx}{\textwidth}{|X|X|X|}
	\hline
	Blazar & Disc Luminosity, $L_d$ ( $\times 10^{46}$ $erg\,  s^{-1}$) & Estimated Disc Luminosity ( $\times 10^{46}$ $erg \,s^{-1}$)\\
	\hline
	\hline
        PKS 0208-512 & 1.47 ($\pm$0.74) & 1.74 ($\pm$0.44)\\
        \hline
        PKS 0454-234 & 1.88 ($\pm$0.94) & 3.33 ($\pm$1.85)\\
        \hline
        3C 273 & 4.8 ($\pm$2.4) & 3.01 ($\pm$0.26)\\
        \hline
        3C 279 & 0.3 ($\pm$0.1) & 0.54 ($\pm$0.28)\\
	\hline
        PKS 2052-474 & 3.75 ($\pm$1.87) & 1.29 ($\pm$1.68)\\
	\hline
        PKS 1510-089 & 0.42 ($\pm$0.21) & 0.34 ($\pm$0.17)\\
        \hline
        3C 454.3 & 3 ($\pm$1.5) & 2.51 ($\pm$0.67)\\
        \hline
	\end{tabularx}
\label{TabDiscLum}
\end{table*}
%
Figure \ref{Bin} exhibits the decomposed disc and jet light curves for a number of blazars in our sample binned over 5-day intervals. The plots of the remaining blazars are shown in Appendix B: Figure \ref{Avg_Remaining} and \ref{2142_Avg}. The ratio of the PL to BB component indicates whether the thermal signature is prominent (low state; BB $>$ PL) or the non-thermal radiation dominates the net flux (high state; PL $\geq$ BB). The low state epochs of the blazars PKS 0208-512 \citep{2013ApJ...771L..25C} and PKS 1510-089 \citep{2020MNRAS.498.5128R} are consistent with the period of low state obtained from decomposing the PL-BB components in our analysis. From this analysis, we get an idea of how the PL and BB components behave over time. For 10 out of 13 blazars in our sample, we find that the former dominates the latter in most epochs, i.e., $\sim$90\% of the observation duration. This agrees with the fact that the jet emission dominates the observed SED of blazars.
\par
We carry out the disc-jet decomposition of the OIR fluxes binned at an interval of 1 day of all the blazars in our sample. Figure \ref{DCCF} shows the decomposed disc and jet light curves of six blazars in our sample as examples. The plots of the remaining blazars are given in Appendix C: Figures \ref{DCCF_Remaining} and \ref{2142_LC}. The disc component is usually high when the jet is high and vice-versa. However, the variability amplitude of the disc emission at a given time interval is smaller than that of the jet. 
3C 454.3 is a blazar in high state through the majority of the duration of the observations used here, as seen in Fig. \ref{DCCF}. For 3C 273 and PKS 1510-089, shown in Fig. \ref{DCCF}, the disc component is prominent at most epochs.
\par

\subsection{Verification I: To Retrieve Given Parameters from Simulated Data}
In order to verify that our employed method is able to obtain the parameters of the power-law and the blackbody components accurately, we simulate data using Eqn. \ref{eqn1},
where we use fixed values of $a$ and $b$, and vary those of $c_b$ and $d$ as given in Table \ref{Tab1}. In our actual data, the power-law component from the jet dominates over the blackbody component, which is often much smaller than the former. We fix $a$ and $b$ at typical values for blazars in our sample, and used a range of values of $c_b$ and $d$ such that the resultant PL/BB ratio spans a large range. $c_b$ is the normalisation constant associated with BB. Thus, a stronger disc will have higher value of $c_b$. $d$ is inversely related to the temperature. A smaller value of $d$ indicates a hotter disc. Hence, we use a range of values of $c_b$ and $d$ to represent weaker to stronger, and hotter to cooler discs. 
Then we fit the simulated data with the model given by Eqn. \ref{eqn1} to check whether we can recover the input values of $a, b, c_b, d$ from the best-fit. 
\par
Table \ref{Tab2} exhibits the recovered parameters $a',b',c_b'$ and $d'$ along with the input values. It shows that the recovered parameters are retrieved accurately for all cases except when the values of both $c_b$ and the temperature are extremely high \{$c_b=17,d=0.2,$ Temperature$=24000K$\}. This corresponds to the case when the blackbody component is much higher than the power-law, which is not applicable to blazar emission in most cases. 


\begin{table*}
	\centering
	\small
	\caption{Parameter Comparison (see text in Section 3.3). The parameters supplied to simulate the data are: $a$ and $c_b$ as normalization constants, $b$ as the spectral index, $d$ as a parameter related to the temperature of the blackbody. $a'$, $b'$, $c_b'$ and $d'$ are the corresponding parameters obtained from our fitting method.}
	\begin{tabularx}{\textwidth}{|X|X||X|X||X|X||X|X|}
	\hline
	a&a'$(\pm0.021)$&b&b'$(\pm0.015)$&$c_b$&$c_b'$$(\pm0.00065)$&d&d'$(\pm0.009)$\\
	\hline
	\hline
	4.179&4.179&1.806&1.806&0.06&0.06&0.2&0.2\\
	\hline
	4.179&4.179&1.806&1.806&0.06&0.06&0.5&0.5\\
	\hline
	4.179&4.179&1.806&1.806&0.06&0.06&0.8&0.8\\
	\hline
	4.179&4.179&1.806&1.806&0.06&0.06&1.8&1.8\\
	\hline
	4.179&4.179&1.806&1.806&0.06&0.06&3.0&3.0\\
	\hline
	\hline
	4.179&4.179&1.806&1.806&0.2&0.2&0.2&0.2\\
	\hline
	4.179&4.179&1.806&1.806&0.2&0.2&0.5&0.5\\
	\hline
	4.179&4.179&1.806&1.806&0.2&0.2&0.8&0.8\\
	\hline
	4.179&4.179&1.806&1.806&0.2&0.2&1.8&1.8\\
	\hline
	4.179&4.179&1.806&1.806&0.2&0.2&3.0&3.0\\
	\hline
	\hline
	4.179&4.179&1.806&1.806&0.8&0.8&0.2&0.2\\
	\hline
	4.179&4.179&1.806&1.806&0.8&0.8&0.5&0.5\\
	\hline
	4.179&4.179&1.806&1.806&0.8&0.8&0.8&0.8\\
	\hline
	4.179&4.179&1.806&1.806&0.8&0.8&1.8&1.8\\
	\hline
	4.179&4.179&1.806&1.806&0.8&0.8&3.0&3.0\\
	\hline
	\hline
	4.179&4.179&1.806&1.806&3.0&3.0&0.2&0.2\\
	\hline
	4.179&4.179&1.806&1.806&3.0&3.0&0.5&0.5\\
	\hline
	4.179&4.179&1.806&1.806&3.0&3.0&0.8&0.8\\
	\hline
	4.179&4.179&1.806&1.806&3.0&3.0&1.8&1.8\\
	\hline
	4.179&4.179&1.806&1.806&3.0&3.0&3.0&3.0\\
	\hline
	\hline
	4.179&4.405e+01&1.806&6.981e+02&17&1.708e+01&0.2&2.005e-01\\
	\hline
	4.179&4.179&1.806&1.806&17.0&17.0&0.5&0.5\\
	\hline
	4.179&4.179&1.806&1.806&17.0&17.0&0.8&0.8\\
	\hline
	4.179&4.179&1.806&1.806&17.0&17.0&1.8&1.8\\
	\hline
	4.179&4.179&1.806&1.806&17.0&17.0&3.0&3.0\\
	\hline
	\end{tabularx}
	\label{Tab2}
\end{table*}
\par

\subsection{Verification II: To Retrieve Constant Disc Component from Variable Flux}
As a second check of our analysis we verify that our model can retrieve the component light curves accurately even when disc emission is constant. The optical variability (e.g., flux doubling) timescale in radio-weak AGN, in which the optical-NIR emission is supposedly from the accretion disk, is typically a few hundred days \citep[e.g., ][]{2005ApJ...622..129S, 2021iSci...24j2557C}. On the other hand, that in low-synchrotron-peaked blazars, in which the optical-NIR emission is dominated by the jet emission, is a few days \citep[e.g., ][]{2019ApJ...880...32Liodakis, 2012ApJ...749..191C, 2019MNRAS.490..124M}. Thus, in blazars, the disc is regarded as fairly stable as compared to the highly variable jet. Hence, it is important to check whether our fitting procedure can extract a constant disc component from a highly variable total flux, especially when the disc component has a considerably lower flux than the jet. Data are simulated using Equation \ref{eqn1}.
\begin{figure*}
\centering
\includegraphics[width=\textwidth]{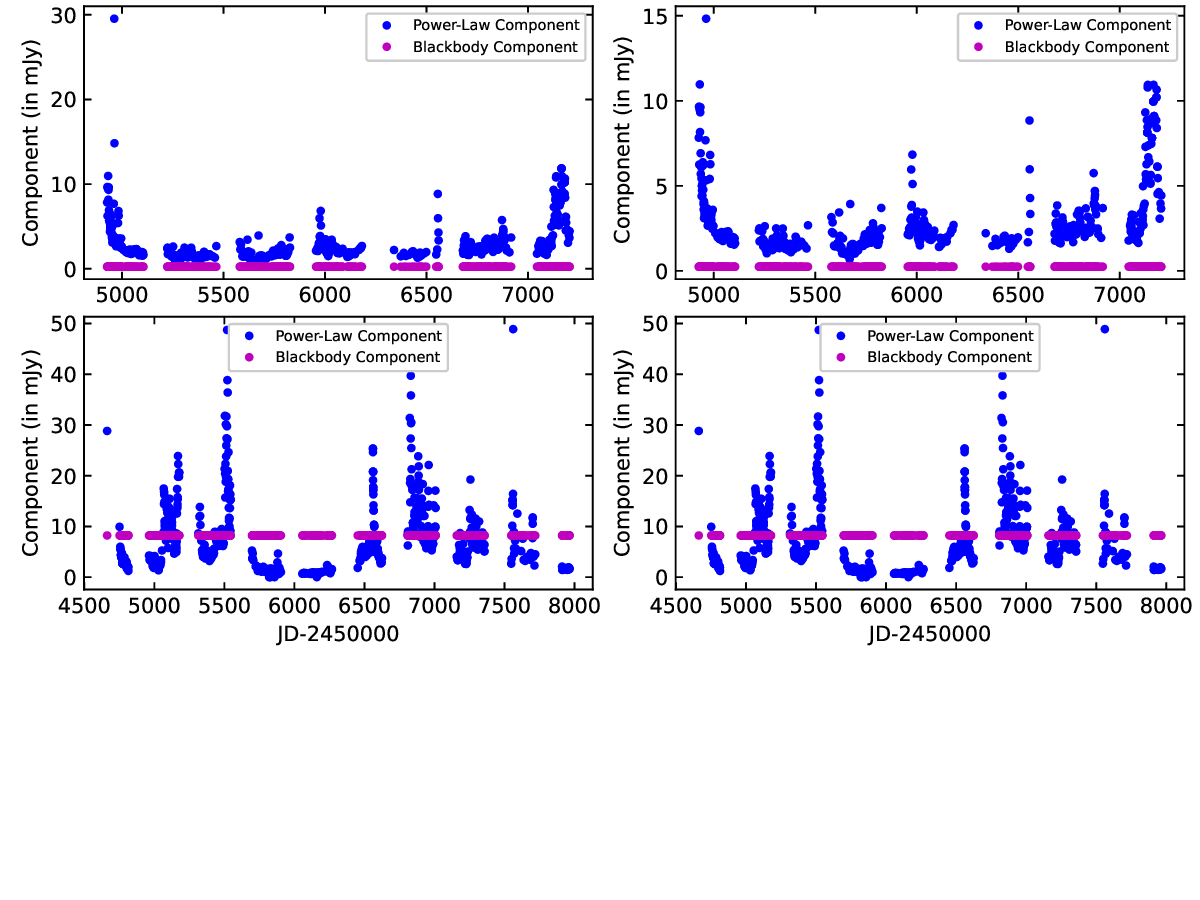}
\caption{\textbf{Top Left:} Simulated light curve of components generated using the PL parameters of PKS 1510-089. \textbf{Top Right:}  Light curve of components extracted from the simulated light curves shown on the left. \textbf{Bottom Left:} Model light curve of components generated using the PL parameters of 3C 454.3. \textbf{Bottom Right:}  Light curve of components extracted from the simulated light curves shown on the left. } 
\label{Model}
\end{figure*}
The value of parameters $a$, $b$, $c_b$ and $d$ were obtained by analysing the light curves of PKS 1510-089 and 3C 454.3. To generate our model data, we utilized $a$ and $b$ to simulate the time-varying PL component. We fixed $c_b$ and $d$ for a constant disc component throughout the time span of the light curve. Thus the disc component of the model data remained fixed throughout the light curve, whereas the jet component varied. For the model data set, using PKS 1510-089 parameters, the fixed disc parameters were $c_b=0.009$ and $d=0.3$. For 3C 454.3, $c_b=0.3$ and $d=0.3$ were used. The latter was used as an upper limit, since the disc cannot usually be seen to outshine the jet by such large amount. The four new parameters $a$, $b$, $c_b$ and $d$ were plugged in Eqn. \ref{eqn1} for five values of total fluxes, corresponding to the five $BVRJK$ bands for each day, thus generating the model light curves.
\begin{figure*}
\centering
\includegraphics[width=\textwidth]{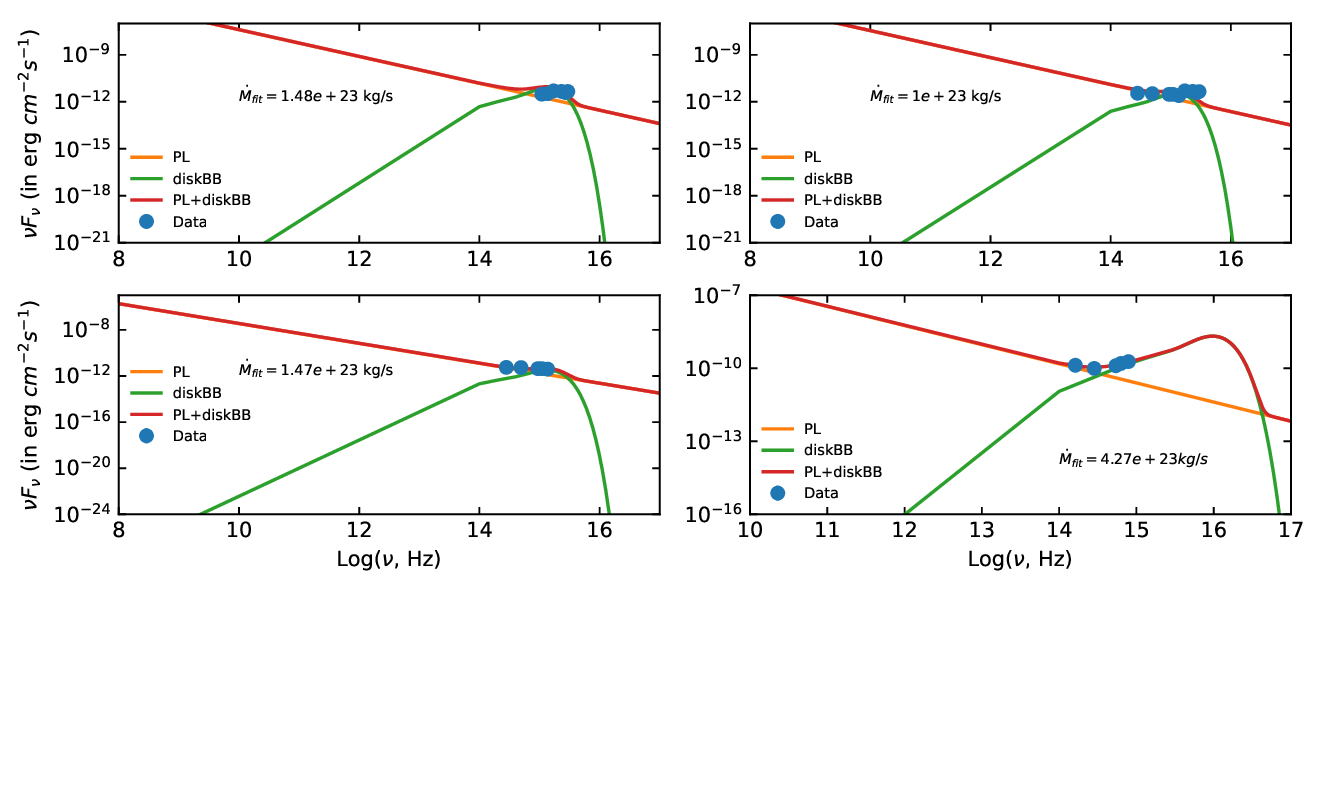}
\caption{Combined power-law and multicolor blackbody curve best fit, in the blazar rest frame, using \textbf{Top Left:} UVOT data of PKS 0208-512. \textbf{Top Right:} SMARTS$+$UVOT data of PKS 0208-512. \textbf{Bottom Left:} SMARTS data of PKS 0208-512. \textbf{Bottom Right:} SMARTS data of 3C 273}
\label{SED}
\end{figure*}
\par
Figure \ref{Model} exhibits the plots resulting from the above analysis. We find that the input and retrieved light curves in Fig. \ref{Model} are identical for both a very low value and a value close to unity of the ratio of the disc to jet flux. Therefore, we conclude that we are able to extract a constant disc and varying jet successfully.
\subsection{Verification III: Deriving Disc Component from OIR and Additional UV Data}
\label{sec3.5}
\subsubsection{Data Analysis}
In this work, we decompose the $BVRJK$-band data into the disc and jet components by fitting with a model consisting of a combination of a simple blackbody and a power-law. However, accretion disc in AGN usually peaks in the UV band. Therefore, including UV data in addition to the $BVRJK$-band will make the analysis more robust. However, simultaneous UV data are not available for those blazars at most of the epochs. Furthermore, attempting to fit a more rigorous model, e.g., replacing the simple blackbody with a multi-color blackbody in the above model, to only five optical-near IR data points for each epoch does not lead to robust estimation of parameters. Therefore, we collected UV data points simultaneous with the SMARTS observations for a blazar and fit those data with a model comprised of a power-law and a multi-color blackbody, in the blazar rest frame. 

For the above purpose, we utilize the UV data points available at the SSDC website{\footnote{\href{https://www.ssdc.asi.it/mmia/index.php?mission=swiftmastr}{https://www.ssdc.asi.it/mmia/index.php?mission=swiftmastr}}} for very few overlapping days. To fit the multi-color blackbody function, we use UV-optical-near IR data, i.e., 4 UV bands - $U, W1, M2,$ and $W2$, 3 optical bands - $B,V,R$ and 2 near IR bands- $J, K$. 
We use the masses of the SMBH provided by \citet{krauss2016tanami} and deduce the disc luminosity.
\par
In order to extract the desired parameters from the data points of a particular epoch, we can use a combination of power-law and multicolor blackbody as our fitting function
\bbe
\nu F_\nu &=& a\nu^{-b} + C_b\times\nu \frac{4h\pi cos (i) \nu^3}{c^2 D_A^2} \int_{R_{in}}^{R_{out}} \frac{ RdR}{e^{h\nu/kT(R)}-1},
\ee
where $a=$ normalization constant of power-law, $b=$ spectral index, $C_b=$ normalization constant of the multicolor blackbody, $\dot M=$ accretion rate. To model the accretion disc, we have used \citep{2013A&A...560A..28C},
\bbe
T^4(R) &=& \frac{3GM\dot M}{8\pi R^3 \sigma}\left[1-\sqrt{\frac{R_{in}}{R}}\right]
\ee
where $T(R)=$ temperature at radius $R$, $M=$ mass of the central SMBH, $\dot M$= accretion rate, $G=$ gravitational constant, $R=$ radius, $\sigma$= Stefan-Boltzmann constant, $\epsilon=$ accretion efficiency, $c=$ speed of light, $L=$ luminosity. $R_{in}=$ inner radius of the disc, $R_{out}=$ the outer radius of the disc, $D_A$ is the angular diameter distance, $i$ is the angle to the line of sight. We use $R_G = \frac{GM}{c^2}$, $R_{in} = 3R_G$, and $R_{out} = 10^4 R_{in}$, where $R_G$ is the gravitational radius. The disc temperature decreases with increase in radius, and provides negligible contribution beyond $R_{out}$. 
\subsubsection{Results}
Figure \ref{SED} gives the plots of the combined multicolor blackbody plus power-law fit to various data sets of the same blazar PKS 0208-512. What we intend to check with the three separate data sets is their agreement in optimizing the desired parameters. In the first case, the UV-optical data points have all been taken from the \textit{Swift-UVOT} database and they are simultaneous. In the second case, we have taken the UV points from \textit{Swift} and $BVRJK$ data points from the Yale-SMARTS blazar database, which are quasi-simultaneous. 
In Case 3, we use only the $BVRJK$ data points of an epoch near the \textit{Swift} data. This is the case we use for most of our analysis with one change, namely, we use multicolor blackbody here instead of a simple blackbody. Table \ref{Tab3} provides the values of the estimated parameters. 
\begin{table*}
	\centering
 \caption{Parameters optimized for PKS 0208-512.}
	\small
	\begin{tabular}{|c|c||c|c|c|c|}
	\hline
	Case & Data & a & b & $C_b$&$\dot M ( kg/s)$\\
	\hline
	\hline
    1&UVOT data &15.1($\pm 6.2$)& 0.857($\pm0.194$)&0.94$(\pm0.66)$&$1.48(\pm0.19)\times 10^{23}$\\
	\hline
    2&SMARTS+UVOT data&14.9($\pm 6.1$)& 0.862($\pm0.194$)& 0.63$(\pm0.44)$& $1.00(\pm0.12)\times 10^{23}$\\
	\hline
    3&$BVRJK$ data (SMARTS)&14.6$(\pm5.9)$& 0.862$(\pm0.195)$& 0.41$(\pm0.28)$& $1.47(\pm0.19)\times 10^{23}$\\
	\hline
	\end{tabular}
	\label{Tab3}
\end{table*}
\par
We can calculate the disc luminosity from the extracted parameter - accretion rate, using $L_{disc} = \epsilon \dot M c^2$ with $\epsilon = 0.1$. Table \ref{Tab4} gives the comparison of our calculated values of $L_{disc}$ with those in the literature \citep{ghisellini2010general}.
\begin{table*}
	\centering
	\small
	\caption{Comparison of disc luminosity (in $erg/s$) of PKS 0208-512 derived in the 3 cases described in Table \ref{Tab3} and Case 4 (which fits a simple blackbody and power law on the data of Case 3).}
	\begin{tabular}{|c|c|c|c|c|}
	\hline
	Literature, $L_d$ $(\times 10^{46})$ \citep{ghisellini2010general}&Case 1 $(\times 10^{46})$& Case 2 $(\times 10^{46})$&Case 3 $(\times 10^{46})$&Case 4 $(\times 10^{46})$\\
	\hline
	1.47($\pm$ 0.74)&1.33($\pm$ 0.17)&0.90$(\pm 0.11)$&1.32($\pm$0.17)&1.74 ($\pm$0.44)\\
	\hline
	\end{tabular}
	\label{Tab4}
\end{table*}
\par
From Table \ref{Tab3}, we see that in all the three cases, there is a good agreement among the optimized parameters. This suggests that the parameters derived from curve fitting on simultaneous data points may be replicated by quasi-simultaneous optical-UV data points. This can be reconciled with the fact that the disc, which peaks in the optical-UV regime, does not vary greatly on $\sim$days-weeks timescale. In Table \ref{Tab4}, we add another case, denoted by `case 4,' which is the same SMARTS data as case 3 but fitted with Eqn. \ref{eqn1} (simple blackbody + power law). Again, we find that the result is consistent with the other cases and literature. It is important to note that case 4, using the general fitting method of this paper, gives a disc luminosity which reconciles within the errors with the disc luminosity given in literature \citep{ghisellini2010general}. \citet{ghisellini2010general} obtained the disc luminosity by modeling the broadband SED of the source. Whereas we have obtained the same, within the uncertainties, by fitting the OIR data with a multicolor blackbody as well as a simple blackbody function. Assuming the disc luminosity does not change beyond the range of uncertainties ($\sim$50\%) given in \citet{ghisellini2010general} between the epochs of observations of those data sets, the above consistency of the results provides further confidence in our method. This was also seen in Table \ref{TabTemp} and \ref{TabDiscLum}. In addition, \citet{krauss2016tanami} found a simple blackbody described the SEDs of multiple blazars better than a model with multi-colored blackbody. This provides further strength to our previous analysis using a simple blackbody.
\par
On extending our analysis to another blazar, 3C 273 (Fig. \ref{SED}, bottom right panel), we see that the estimated parameter, from multicolor blackbody fit, $L_{disc} = (3.84 \pm 0.13)\times10^{46}$ erg\,s$^{-1}$ 
is consistent with $L_{disc} = (4.8 \pm2.4)\times10^{46}$ erg\,s$^{-1}$ found in the literature \citep{ghisellini2010general}. As shown in Table \ref{TabDiscLum}, using simple blackbody fit on the same SMARTS data, we obtain $L_{disc} = (3.01 \pm 0.26)\times10^{46}$ erg\,s$^{-1}$, which agrees with the literature as well. This further verifies that our method provides reliable results regarding the parameters we extract from it.
\section{Cross-Correlation of the Disc and Jet Components}
\label{sec4}
Figure \ref{DCCF} exhibits the discrete cross-correlation function \citep[DCCF;][]{1988ApJ...333..646E} between the variability of the decomposed disc and jet components in six FSRQs. We find that for most of the cases the correlation time lag between the disc and jet variability is consistent with being less than 10 days. In the rest of the sources, shown in Appendix C: Figures \ref{DCCF_Remaining} and \ref{2142_DCCF}, it is less than 20 days. For very few cases where there are large gaps in the light curves, we get time lag $>$ 20 days, which may be unreliable. 
\par
The relation between accretion disc power, obtained from emission line luminosities, and jet power, obtained from various indicators including unbeamed $\gamma$-ray luminosity, radio luminosity of the large-scale jet and apparent speed of radio knots moving down the pc-scale jet, in a sample of few hundred blazars was studied by \citet{2022PhRvD.106f3001R}. They showed that the disc and jet power are correlated although with considerable scatter. In this work, we have studied the correlation between the temporal variation of the disc and jet power in the blazars in our sample. We find a similar trend that the variations are correlated in most of the cases although the peak of the correlation function is not always very high indicating the correlation is moderate.
\section{Discussion and Conclusion}
\label{sec5}
\par
We use a combination of a power-law and a blackbody function to fit the OIR spectra of a sample of 13 blazars using light curves spanning 8 years. From the best-fit parameters we obtain the disc and jet components contributing to the total flux, in the blazar rest frame. OIR emission of blazars are often dominated by the jet component and the contribution of the disc may be negligibly small. Therefore, in order to verify that the disc-jet decomposition of the OIR flux that we are carrying out is accurate we test our method using simulated data with various relative contributions of the disc and jet, and find that we recover the correct values with small relative uncertainties. The disc emission is more prominent in the UV wave band. We use a more detailed model to fit the OIR data for two blazars, along with (quasi)simultaneous UV data points, in which such data are available, to check that the results we obtain are consistent with that derived using our model on the OIR data. The above tests ensure that the disc contribution to the total OIR flux that we derive using our method is accurate even when it is small compared to the jet component. In order to study the variability we computed the discrete cross-correlation function and found that the disc and jet light curves closely follow each other.

Blazars, having a prominent jet, are suitable sources to study the launching and collimation of relativistic jets. Connection of the emission and dynamics of the jet with the state of the accretion disc is a useful avenue to probe the above. However, the disc emission in blazars is often overwhelmed by the beamed emission from the jet. Therefore, disc emission in blazars is identified using some indirect methods, such as, through its relation with emission line luminosities \citep[e.g.,][]{2021ApJS..253...46P}, by eliminating the jet contribution estimated from its $\gamma$-ray luminosity \citep[e.g.,][]{rak20,pan22}, in sources that have a slightly misaligned jet \citep[e.g.,][]{chat09,chat11}. In this paper, we have decomposed the total OIR emission directly into a non-thermal jet component, and a thermal component that is interpreted as that from an accretion disc. Given that $BVRJK$ photometric information, such as used here, are available for many blazars, this method may be useful to estimate the disc and jet contributions without using other indirect indicators.


\section*{Data Availability}
\label{sec6}
The data utilized in this work have been taken from:-
\begin{enumerate}
\item This paper has made use of up-to-date SMARTS optical/near-infrared light curves \citep{2012ApJ...756...13B} that are available at weblink: \url{http://www.astro.yale.edu/smarts/glast/home.php}
\item NASA/IPAC Extragalactic Database. Weblink: \url{https://ned.ipac.caltech.edu/}
\item Space Science Data Center. Weblink: \url{https://www.ssdc.asi.it/mmia/index.php?mission=swiftmastr}
\end{enumerate}


\section*{ACKNOWLEDGEMENTS}
We thank the anonymous referee for useful comments which helped us improve the manuscript. GR acknowledges the DST INSPIRE Scholarship and JBNSTS Scholarship. RC thanks Presidency University for support under the Faculty Research and Professional Development (FRPDF) Grant, ISRO for support under the AstroSat archival data utilization program, DST SERB for a SURE grant (File No. SUR/2022/001503), and IUCAA for their hospitality and usage of their facilities during his stay at different times as part of the university associateship program. RC acknowledges financial support from BRNS through a project grant (sanction no: 57/14/10/2019-BRNS) and thanks the project coordinator Pratik Majumdar for support regarding the BRNS project. RC thanks Saumyadip Samui for discussion related to statistical analyses used in this work. RC thanks Charles Bailyn and Meg Urry for useful discussions in which the core idea of this work was developed. 



\bibliographystyle{mnras}
\bibliography{Project1} 



\appendix

\section{Disc-Jet Components of Blazars}
\begin{figure*}
\centering
\includegraphics[width=\textwidth]{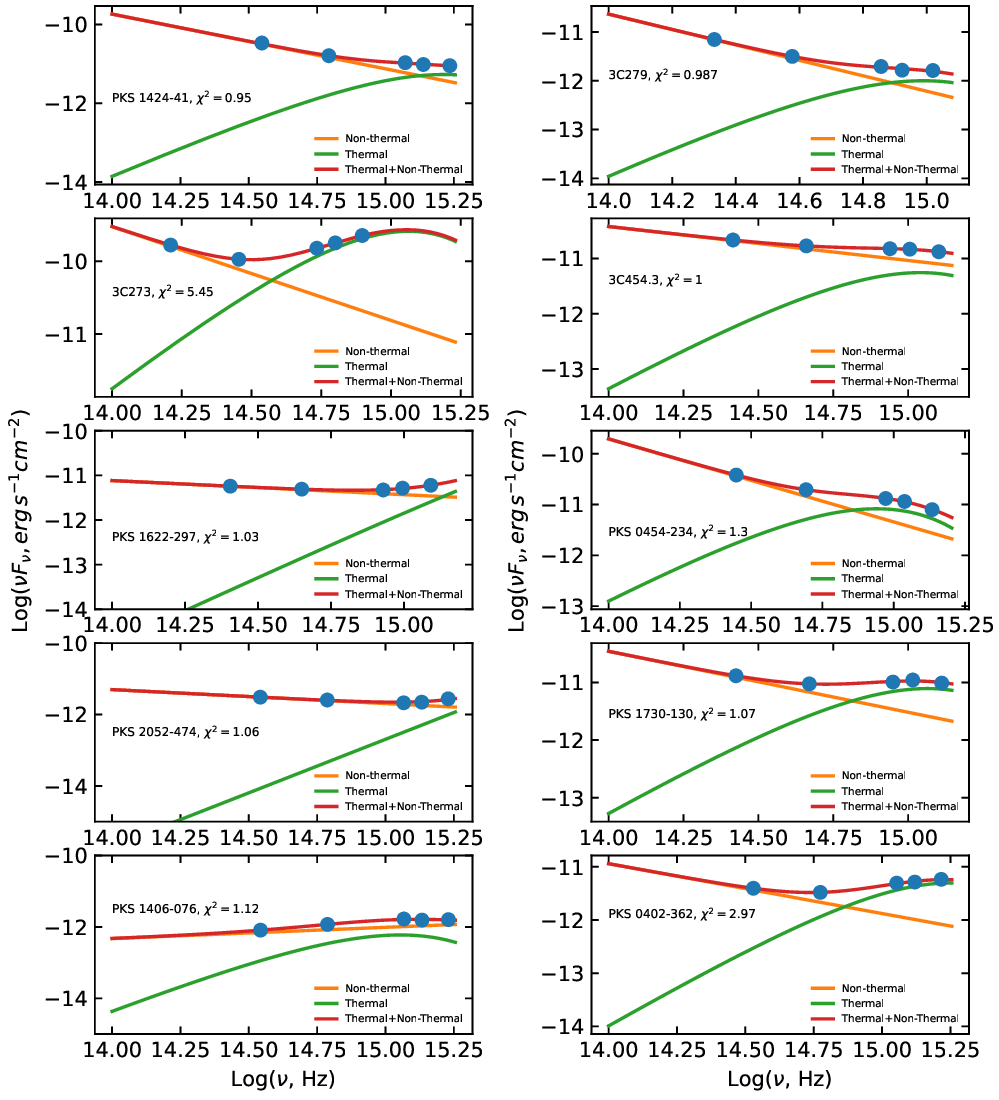}
\caption{Each panel shows the decomposed disc and jet component, in the blazar rest frame, such as that in Fig. \ref{HighLow}. The remaining blazars of our sample are shown in this plot and labelled.}
\label{Fit_Remaining}
\end{figure*}

\section{Average Disc-Jet Components}
\begin{figure*}
\centering
\includegraphics[width=\textwidth]{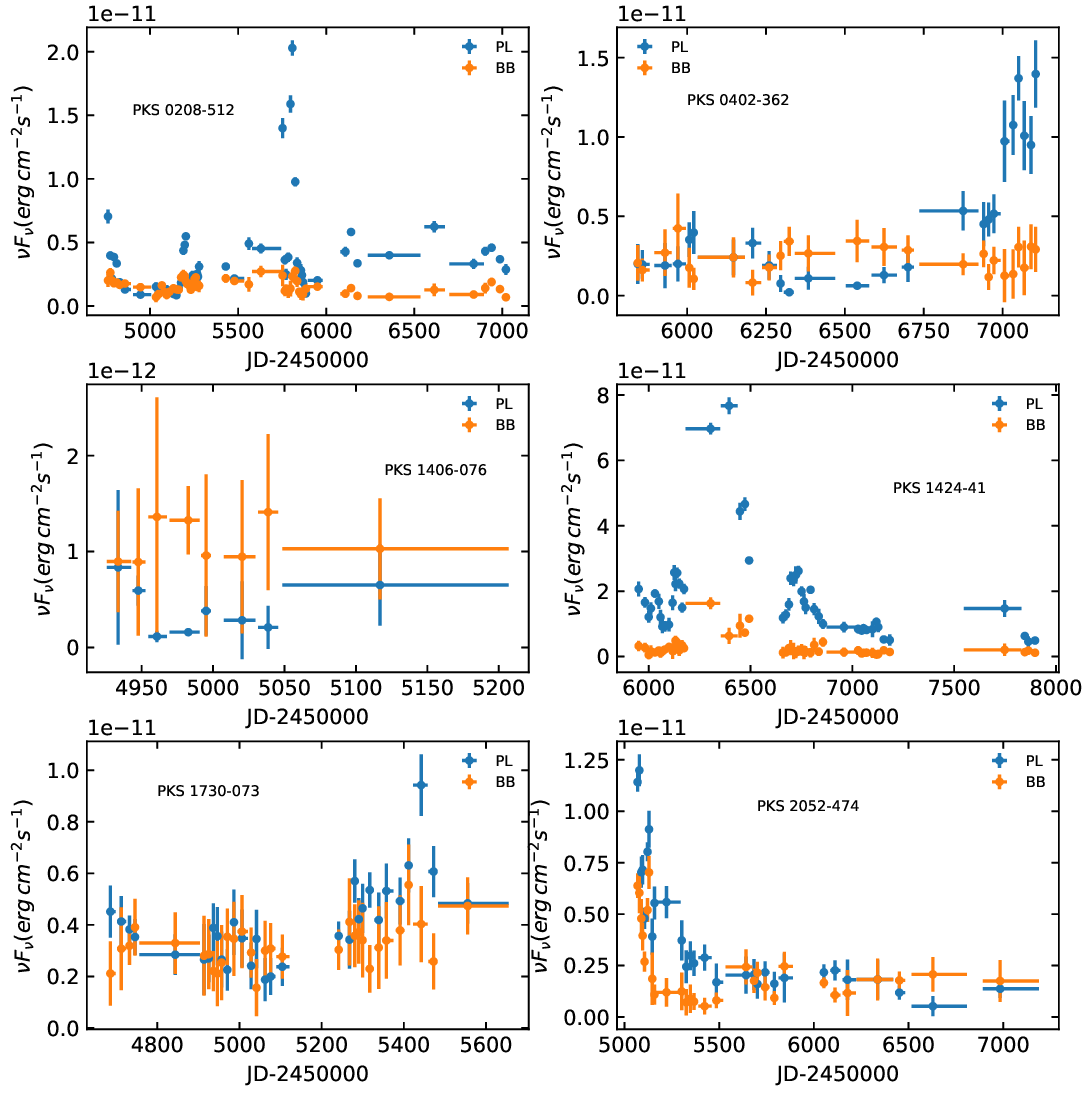}
\caption{Same as Fig. \ref{Bin}, each panel shows the decomposed disc and jet component light curves binned over 5 consecutive days for: \textbf{Top Left:} PKS 0208-512. \textbf{Top Right:} PKS 0402-362.  \textbf{Middle Left:} PKS 1406-076. \textbf{Middle Right:} PKS 1424-41. \textbf{Bottom Left:} PKS 1730-130. \textbf{Bottom Right:} PKS 2052-474.}
\label{Avg_Remaining}
\end{figure*}

\begin{figure}
\centering
\includegraphics[width=8cm,height=6cm]{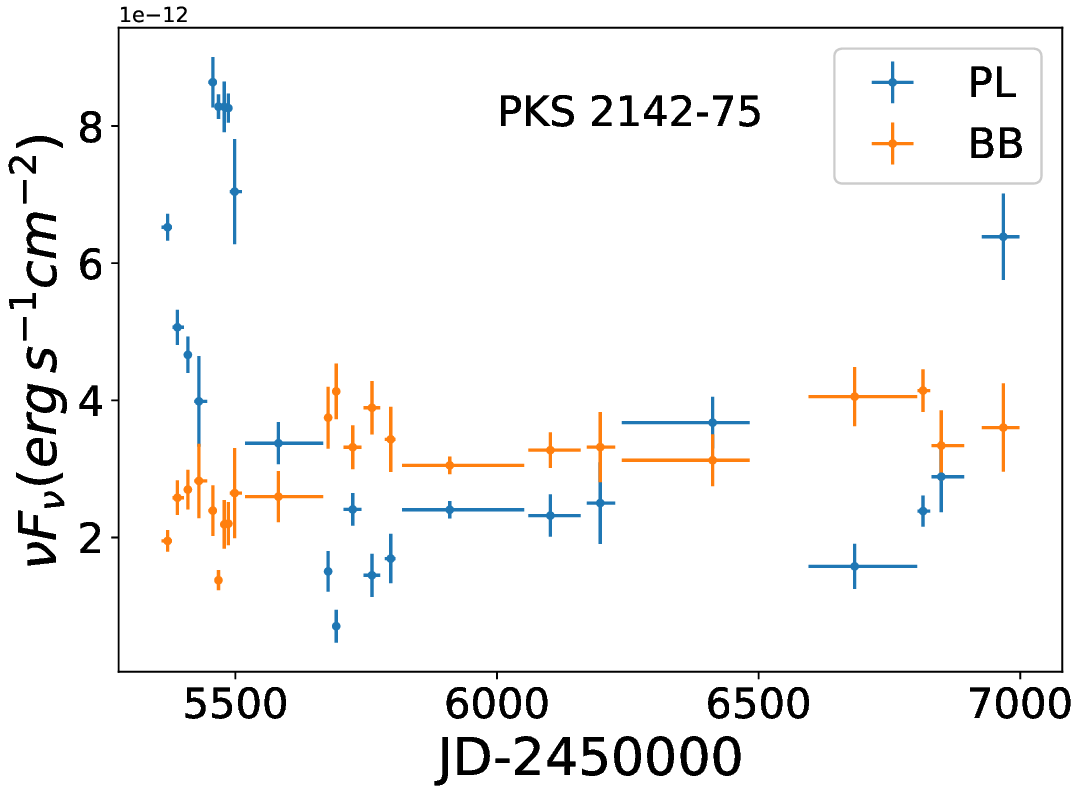}
\caption{The decomposed disc and jet component light curves binned over 5 consecutive days of PKS 2142-75.}
\label{2142_Avg}
\end{figure}

\section{DCCF and Decomposed Disc-Jet Components}
\begin{figure*}
\centering
\includegraphics[width=\textwidth]{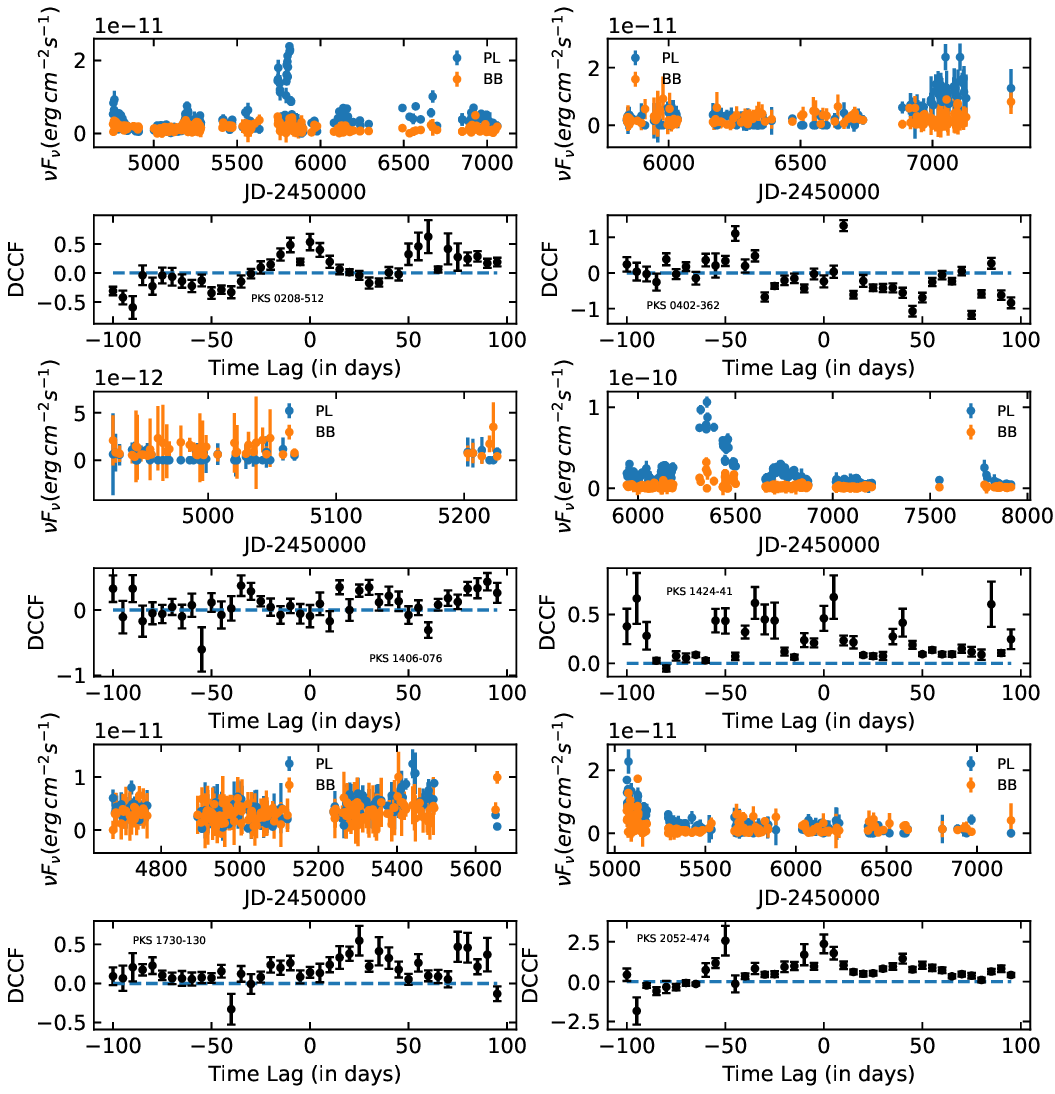}
\caption{Same as Fig. \ref{DCCF}, each panel shows the decomposed disc and jet component light curves and the associated DCCF plot below it. The remaining blazars of our sample are shown in this plot and labelled. \textbf{Top Left:} PKS 0208-512. \textbf{Top Right:} PKS 0402-362.  \textbf{Middle Left:} PKS 1406-076. \textbf{Middle Right:} PKS 1424-41. \textbf{Bottom Left:} PKS 1730-130. \textbf{Bottom Right:} PKS 2052-474.}
\label{DCCF_Remaining}
\end{figure*}

\begin{figure}
\centering
\includegraphics[width=8cm,height=6cm]{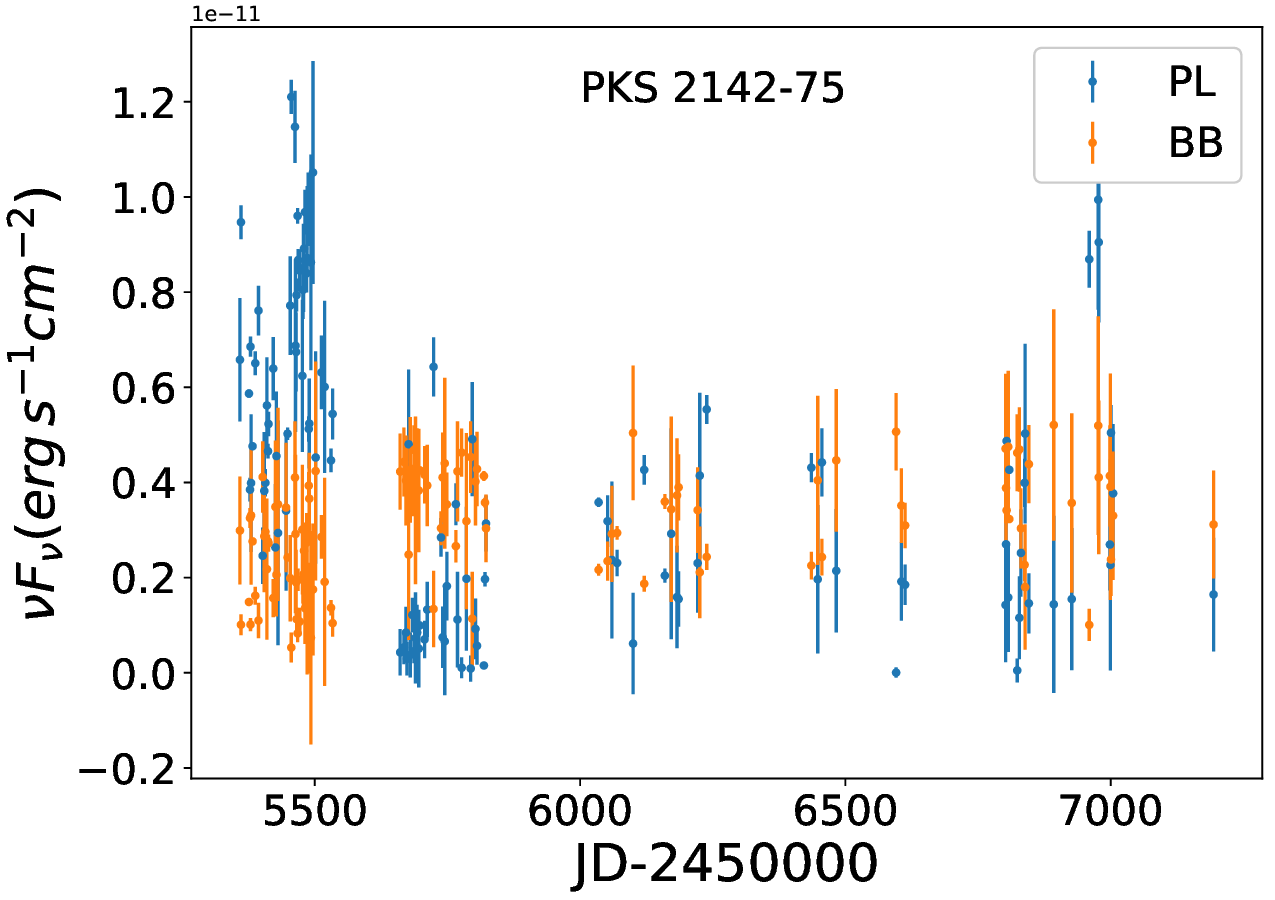}
\caption{The decomposed disc and jet component light curves of PKS 2142-75.}
\label{2142_LC}
\end{figure}

\begin{figure}
\centering
\includegraphics[width=8cm,height=6cm]{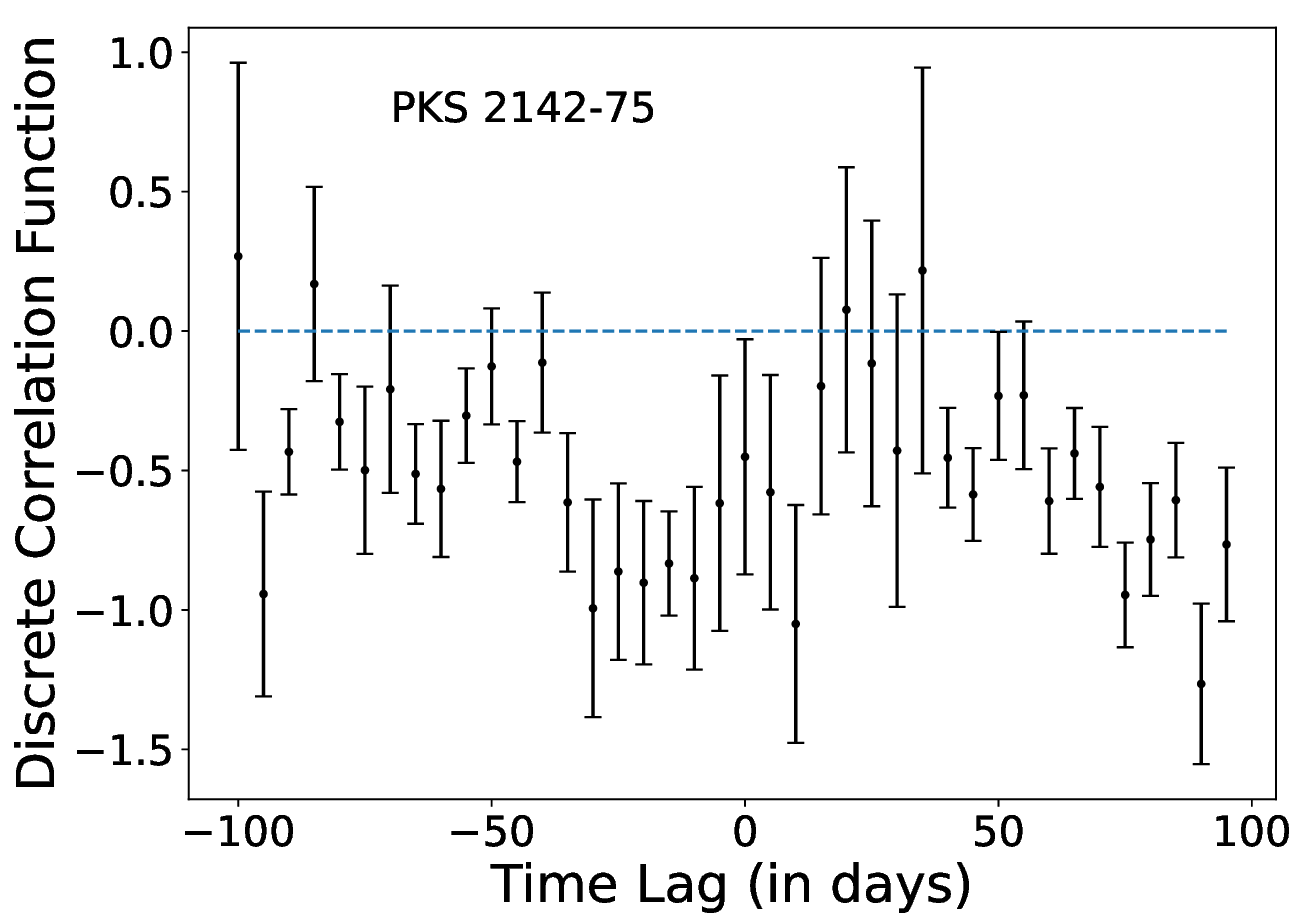}
\caption{DCCF of the decomposed disc and jet component light curves of PKS 2142-75.}
\label{2142_DCCF}
\end{figure}


\bsp	
\label{lastpage}
\end{document}